\documentclass[preprint,12pt,a4paper]{elsarticle}

\usepackage{amssymb}
\usepackage{xcolor}

\journal{J. Mag. Mag. Mater.}

\begin{document}

\begin{frontmatter}

\title{Experimental investigation of magnetic properties of MnFeCo$_{4}$Si$_{2}$ discovered by GNoME}

\author{Shuhei Naganuma}
\address{Department of Electrical Engineering, Faculty of Engineering, Fukuoka Institute of Technology, 3-30-1 Wajiro-higashi, Higashi-ku, Fukuoka 811-0295, Japan}
\author{Jiro Kitagawa}
\address{Department of Electrical Engineering, Faculty of Engineering, Fukuoka Institute of Technology, 3-30-1 Wajiro-higashi, Higashi-ku, Fukuoka 811-0295, Japan}
\ead{j-kitagawa@fit.ac.jp}

\begin{abstract}
AI-driven inorganic materials research has garnered significant attention due to its ability to reduce the time, labor, and cost associated with experiments. An AI model known as GNoME, recently developed by Google DeepMind, is particularly fascinating because it is integrated with the Materials Project open database. The experimental verification of compounds identified by GNoME is a crucial process for advancing AI-driven materials research. Here, we focus on the magnetic compound MnFeCo$_{4}$Si$_{2}$ (Materials ID: mp-3203253), which possesses a layered-like structure. Consistent with the GNoME prediction, MnFeCo$_{4}$Si$_{2}$ crystallizes in a rhombohedral structure with a single-phase nature. We have characterized its magnetic properties and determined that MnFeCo$_{4}$Si$_{2}$ is a soft ferromagnet with a Curie temperature of 1039 K.  
\end{abstract}


\end{frontmatter}

\clearpage

\section{Introduction}
In materials research, traditional materials design approaches are time-intensive and labor-intensive processes. 
However, AI technologies are now transforming the materials research workflow and have the potential to reduce the time, labor, and cost of experiments\cite{Stanev:CM2021,Choudhary:NCM2022}. 
Focusing on inorganic compounds, magnetic materials, energy materials, structural materials, and others have been discovered through AI-driven materials design\cite{Castro:NAM:2020,Rao:Science2022,Hou:AAMI2019,Jiang:JMST2022}. 
Recently, the research team at Google DeepMind reported an AI model based on graph networks for materials exploration (GNoME)\cite{Merchant:Nature2023}.
GNoME employs active learning with graph neural networks (GNNs) that model atomic connections within crystalline structures.
The outputs of the GNNs are rigorously evaluated using density functional theory calculations to assess thermodynamic stability. 
GNoME then incorporates these results back into its training data. 
This iterative process improves the reliability of materials prediction.
Importantly, this system is linked to the Materials Project open database\cite{Jain:APLM2013}, allowing easy access to crystal structure information for predicted inorganic compounds. 
At the present stage, 45,608 inorganic compounds are predicted to be thermodynamically stable. 
Thus, the experimental verification of predicted compounds is an indispensable process in AI-driven materials research.

Our team is researching rare-earth-free magnetic compounds. 
This research is crucial for mitigating the issue of the uneven global distribution of rare-earth ores used in the manufacture of rare-earth permanent magnets. 
We have discovered new compounds, such as B-doped Mn-based room-temperature ferromagnets and Co-based geometrically frustrated compounds exhibiting giant low-temperature coercivity\cite{Kitagawa:JMMM2018,Kitagawa:MRX2023}. 
We are currently exploring rare-earth-free magnetic compounds with unique crystal structures, such as those with geometrically frustrated or anisotropic layered structures. 
Surveying the GNoME database in the Materials Project, we developed an interest in MnFeCo$_{4}$Si$_{2}$. 

\begin{figure}
\begin{center}
\includegraphics[width=12cm]{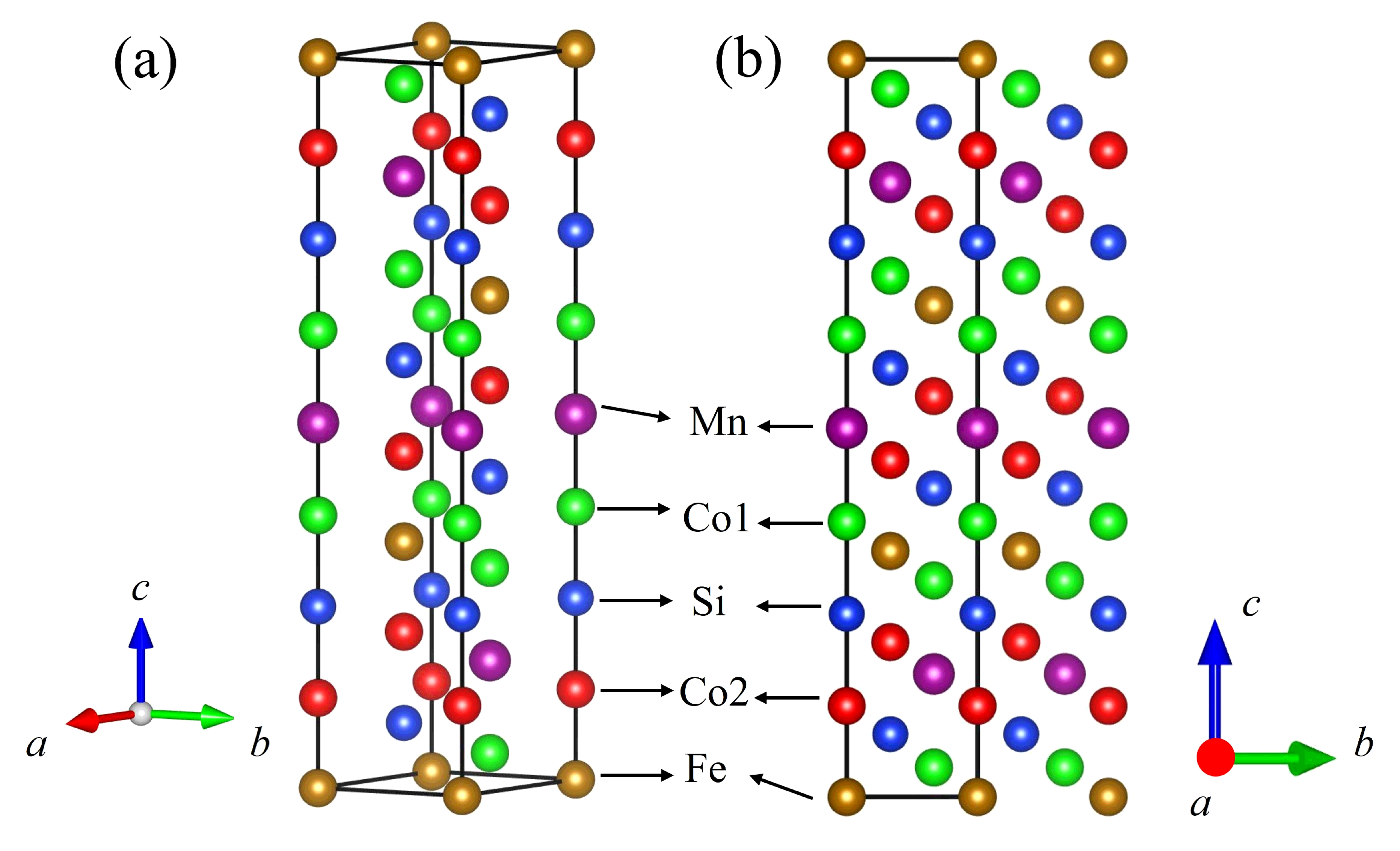}
\end{center}
\caption{Crystal structure of MnFeCo$_{4}$Si$_{2}$ in hexagonal setting. The solid black lines denote the unit cell.}
\label{cry}
\end{figure}

GNoME predicts that MnFeCo$_{4}$Si$_{2}$ (Materials ID: mp-3203253) crystallizes in a rhombohedral structure with the space group {\it R}$\bar{3}${\it m} (No.166).
In the hexagonal setting, the crystal structure is highly anisotropic, with lattice parameters $a$=3.95 \AA \hspace{1mm} and $c$=19.36 \AA \hspace{1mm} (see Fig.\hspace{1mm}\ref{cry}(a)). 
Mn, Fe, and Si atoms occupy Wyckoff positions 3b (0, 0, 0.5), 3a (0, 0, 0), and 6c (0, 0, 0.24831), respectively. 
For Co atoms, there are two 6c Wyckoff sites with atomic positions (0, 0, 0.37328) and (0, 0, 0.12334). 
We refer to Co atoms at (0, 0, 0.37328) and (0, 0, 0.12334) as Co1 and Co2, respectively. 
The crystal structure can be regarded as a layered-like structure with the stacking sequence Co1–Fe–Co1–Si–Co2–Mn–Co2–Si along the c-axis, as shown in Fig.\hspace{1mm}\ref{cry}(b).

In this work, we have synthesized MnFeCo$_{4}$Si$_{2}$ and verified its existence through X-ray diffraction analysis and metallographic examination. 
The magnetic properties of MnFeCo$_{4}$Si$_{2}$ are characterized by magnetization measurements, and several magnetic quantities are compared with those obtained using an electronic-structure calculation software package.

\section{Materials and methods}
The polycrystalline sample ($\sim$2.5 g) was prepared using a custom-built arc furnace. 
The elemental components Mn chips (Kojundo Chemical Laboratory, 99.9 \%), Fe chips (Kojundo Chemical Laboratory, 99.9 \%), Co chips (Rare Metallic, 99.9 \%), and Si chips (Soekawa Rikagaku, 99.999 \%) were employed. 
Owing to the volatile nature of Mn, excess Mn (1.25 at.\%) beyond the stoichiometric composition was added to compensate for weight loss during arc melting. 
Here, we comment on the Mn excess ratio. 
The excess ratio was empirically determined through materials research on Mn-containing compounds. 
In fact, the weight loss during arc melting was approximately 1.1 \%. 
Assuming that the weight loss is attributable to the volatilization of Mn, the chemical composition after arc melting is evaluated to be stoichiometric. 
We note that the excess Mn ratio depends on the performance of the arc furnace and is not directly applicable to furnaces other than ours.
The resulting ingot was sealed in a quartz tube and annealed at 800 $^{\circ}$C for 4 days. 
This annealing condition was selected based on previous studies reporting that compounds containing Fe, Co, or Mn prepared under the same annealing conditions exhibit single-phase formation and sharp magnetic transitions\cite{Kitagawa:JMMM2018,Kitagawa:JMMM2022}.

Room-temperature X-ray diffraction (XRD) patterns were obtained using an X-ray diffractometer (MiniFlex600-C, Rigaku) with Cu-K$\alpha$ radiation. 
Scanning electron microscopy (SEM) images were acquired with a field-emission scanning electron microscope (FE-SEM, JSM-7100F, JEOL), and the chemical composition was evaluated using an energy-dispersive X-ray (EDX) spectrometer integrated with the FE-SEM.

The temperature dependence of dc magnetization $\chi_\mathrm{dc}$ ($T$) was measured using the VersaLab apparatus (Quantum Design) from 50 to 400 K. 
Isothermal magnetization ($M$) curves were obtained with the same equipment. 
To estimate the Curie temperature, $\chi_\mathrm{dc}$ ($T$) measurements were extended up to 1100 K using a vibrating sample magnetometer (TM-VSM33483-HGC, Tamakawa).
Isothermal magnetization curves over the temperature range 650–1095 K were measured using another vibrating sample magnetometer (VSM-5, Toei Industry). 

Electronic-structure calculations were carried out using the Akai-KKR program package\cite{Akai:JPSJ1982}, which is based on the Korringa–Kohn–Rostoker (KKR) method with Green's function formalism. 
The code uses the atomic sphere approximation together with the coherent potential approximation (CPA).
The Perdew–Burke–Ernzerhof (PBE) exchange–correlation potential was adopted, and both spin polarization and spin–orbit coupling were taken into account. 
For Brillouin zone sampling, 690 $k$-points in the irreducible Brillouin zone were used. 
The scattering was considered up to $d$ scattering ($l_\mathrm{max}$ = 2). 

\begin{figure}
\begin{center}
\includegraphics[width=10cm]{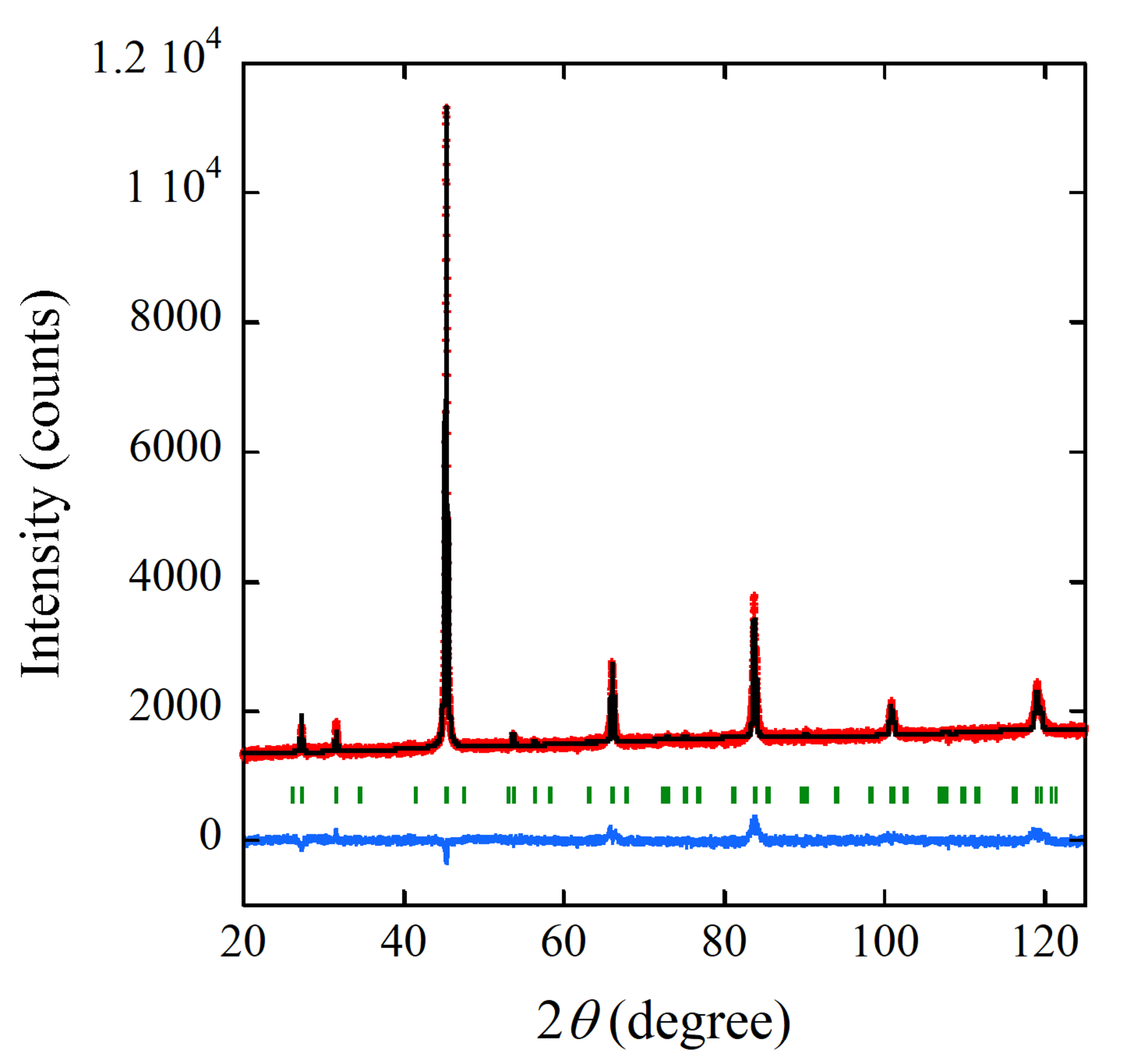}
\end{center}
\caption{XRD pattern of MnFeCo$_{4}$Si$_{2}$. The observed (+) and calculated (solid line) patterns are shown at the top, while the difference between them is displayed at the bottom. Tick marks indicate the positions of Bragg reflections for MnFeCo$_{4}$Si$_{2}$.}
\label{XRD}
\end{figure}

\section{Results and discussion}
\begin{figure}
\begin{center}
\includegraphics[width=8cm]{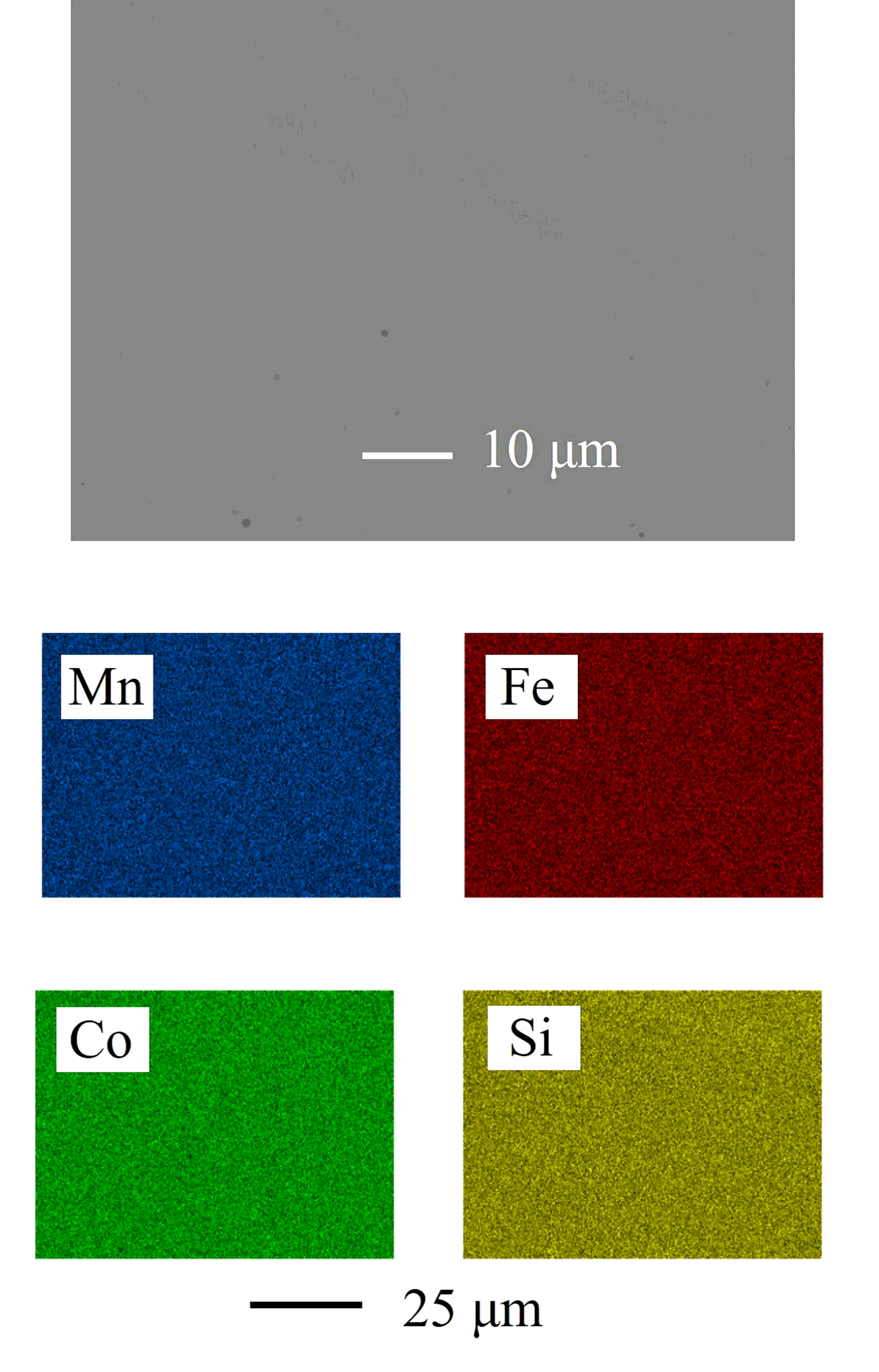}
\end{center}
\caption{SEM image of MnFeCo$_{4}$Si$_{2}$. The elemental mappings of Mn, Fe, Co, and Si are also shown.}
\label{SEM}
\end{figure}

The XRD pattern of MnFeCo$_{4}$Si$_{2}$ is shown in Fig.\hspace{1mm}\ref{XRD} together with the fitted pattern obtained via the Rietveld refinement program\cite{Izumi:SSP2007,Tsubota:SR2017}. 
The good profile fitting supports the thermodynamic stability of the crystal structure proposed by GNoME. 
The lattice parameters in the hexagonal setting are determined to be $a$=3.9978(3) \AA \hspace{1mm} and $c$=19.583(5) \AA, which are slightly larger than the predicted values. 
The crystallite size $D$ was evaluated using the Scherrer equation $D=K\lambda/\beta\mathrm{cos}\theta$, where $\beta$ represents the full width at half maximum of the diffraction peak, $\lambda$ denotes the X-ray wavelength, and $K$ is a numerical coefficient. 
Assuming $K$=0.9, the $D$ value calculated from the main peak is 35 nm.
Figure \ref{SEM} presents the SEM image of MnFeCo$_{4}$Si$_{2}$, revealing no discernible impurity phases. 
The homogeneous elemental mappings in the same figure further support the single-phase nature of the prepared sample. 
The atomic composition determined by EDX analysis is Mn$_{14.0(3)}$Fe$_{12.9(2)}$Co$_{49.9(2)}$Si$_{23.3(4)}$, which is close to the ideal composition Mn$_{12.5}$Fe$_{12.5}$Co$_{50}$Si$_{25}$.
In the SEM image of the present sample without chemical etching, grain boundaries are not observed, as has also been reported for several single-phase alloys and compounds examined without chemical etching\cite{Mohamad:JNST2018,Kitagawa:JALCOM2022,Sarr:JALCOM2025}.
Although the grain size cannot be directly evaluated, the crystallite size $D$ provides a lower bound for the grain size.

\begin{figure}
\begin{center}
\includegraphics[width=14cm]{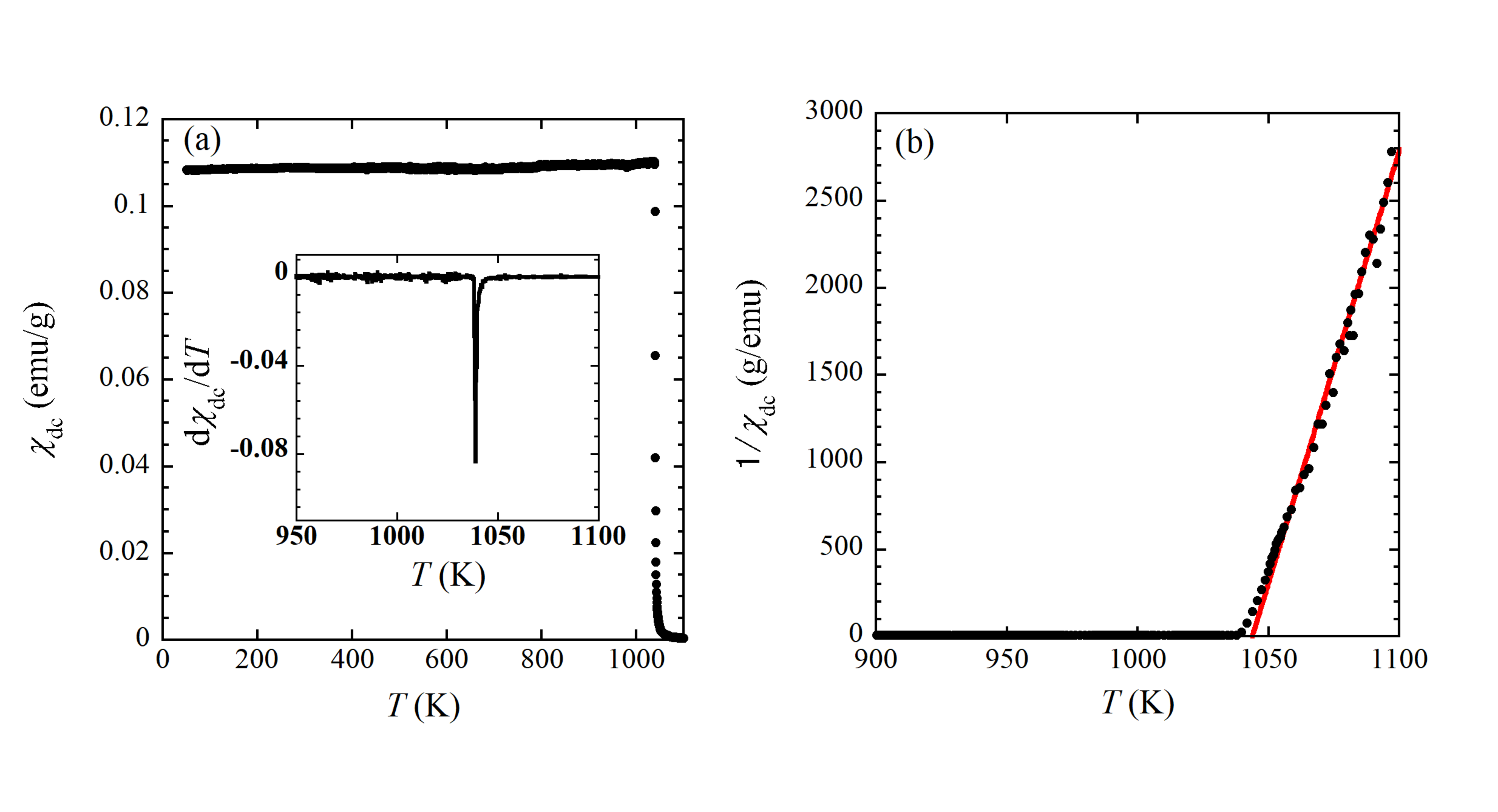}
\end{center}
\caption{(a) Temperature dependent $\chi_\mathrm{dc}$ of MnFeCo$_{4}$Si$_{2}$ measured under external field of 100 Oe. The inset shows the temperature derivative of $\chi_\mathrm{dc}$. (b) Temperature dependence of 1/$\chi_\mathrm{dc}$ for MnFeCo$_{4}$Si$_{2}$. The solid line represents the fit according to the Curie-Weiss law.}
\label{chidc}
\end{figure}

Temperature-dependent $\chi_\mathrm{dc}$ exhibits a rapid increase below approximately 1050 K, indicating the onset of ferromagnetic ordering (Fig.\hspace{1mm}\ref{chidc}(a)). 
The Curie temperature $T_\mathrm{C}$ is determined from the minimum of the temperature derivative of $\chi_\mathrm{dc}$ ($d\chi_\mathrm{dc}/dT$), as commonly employed for many transition-metal-based ferromagnets\cite{Kitagawa:JMMM2018,Kitagawa:MRX2023,Oikawa:APL2001,Yu:APL2003}, and is also shown in Fig.\hspace{1mm}\ref{chidc}(a). 
$d\chi_\mathrm{dc}/dT$ displays a sharp minimum at $T_\mathrm{C}$ = 1039 K. 
As illustrated by the red line in Fig.\hspace{1mm}\ref{chidc}(b), 1/$\chi_\mathrm{dc}$ ($T$) above 1050 K can be described by the Curie–Weiss law:
\begin{equation}
1/\chi_\mathrm{dc}(T)=T/C-\Theta_\mathrm{CW}/C,
\label{equ:RT}
\end{equation}
where $C$ is the Curie constant and $\Theta_\mathrm{CW}$ is the Weiss temperature. 
The effective magnetic moment $\mu_\mathrm{eff}$ derived from $C$ is 3.29 $\mu_\mathrm{B}$ per magnetic ion. 
The positive $\Theta_\mathrm{CW}$ of 1044 K indicates ferromagnetic interactions between magnetic moments and agrees well with the value of $T_\mathrm{C}$.

\begin{figure}
\begin{center}
\includegraphics[width=14cm]{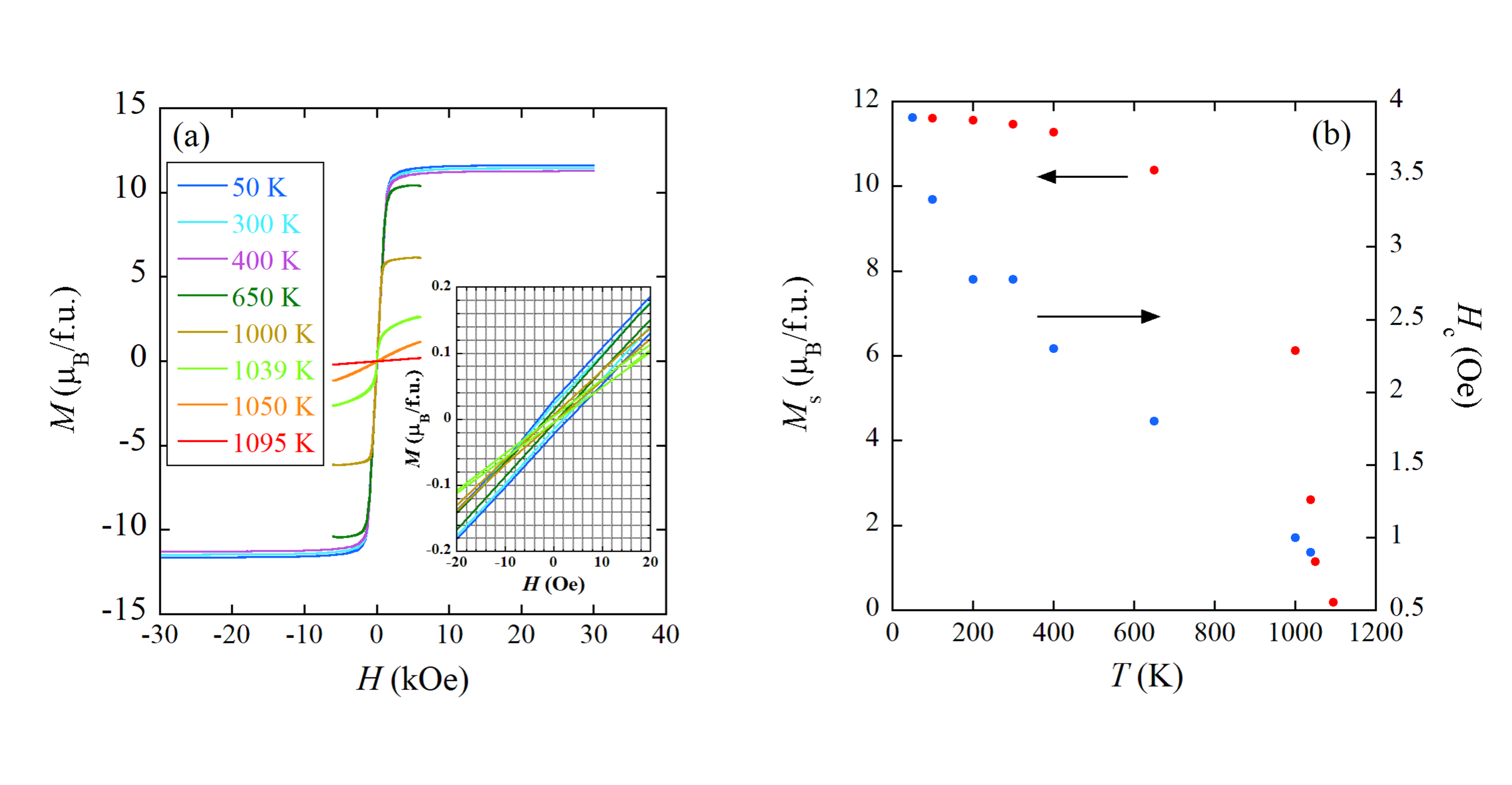}
\end{center}
\caption{(a) Isothermal magnetization curves of MnFeCo$_{4}$Si$_{2}$ measured at 50, 300, 400, 650, 1000, 1039, 1050, and 1095 K. The inset is an expanded view of the low-field region for several curves. (b) Temperature dependence of $M_\mathrm{s}$ and $H_\mathrm{c}$ for MnFeCo$_{4}$Si$_{2}$. The $M_\mathrm{s}$ values were determined from $M$ at 30 kOe for 50 - 400 K and at 6 kOe for 650 - 1095 K.}
\label{MH}
\end{figure}

The isothermal magnetization curves measured over the temperature range of 50–1095 K are shown in Fig.\hspace{1mm}\ref{MH}(a). 
As anticipated from the $\chi_\mathrm{dc}$($T$) behavior, the curves below $T_\mathrm{C}$ display characteristic signatures of ferromagnetism. 
At each temperature below $T_\mathrm{C}$, $M$ rapidly approaches saturation with increasing external field $H$. 
The saturation magnetization $M_\mathrm{s}$ reaches 161.35 emu/g (11.63 $\mu_\mathrm{B}$/f.u.) at 50 K, which is a relatively large value. 
The inset of Fig.\hspace{1mm}\ref{MH}(a) presents an expanded view of several $M$–$H$ curves in the low-field region. 
The extracted coercive fields $H_\mathrm{c}$ are plotted as a function of temperature in Fig.\hspace{1mm}\ref{MH}(b), along with the temperature dependence of $M_\mathrm{s}$, which exhibits the typical behavior of ferromagnets. 
The small value of $H_\mathrm{c}$ classifies MnFeCo$_{4}$Si$_{2}$ as a soft ferromagnetic compound.
$H_\mathrm{c}$ decreases with increasing temperature, which can be ascribed to the enhanced thermal energy at elevated temperatures that facilitates domain-wall motion.

\begin{figure}
\begin{center}
\includegraphics[width=12cm]{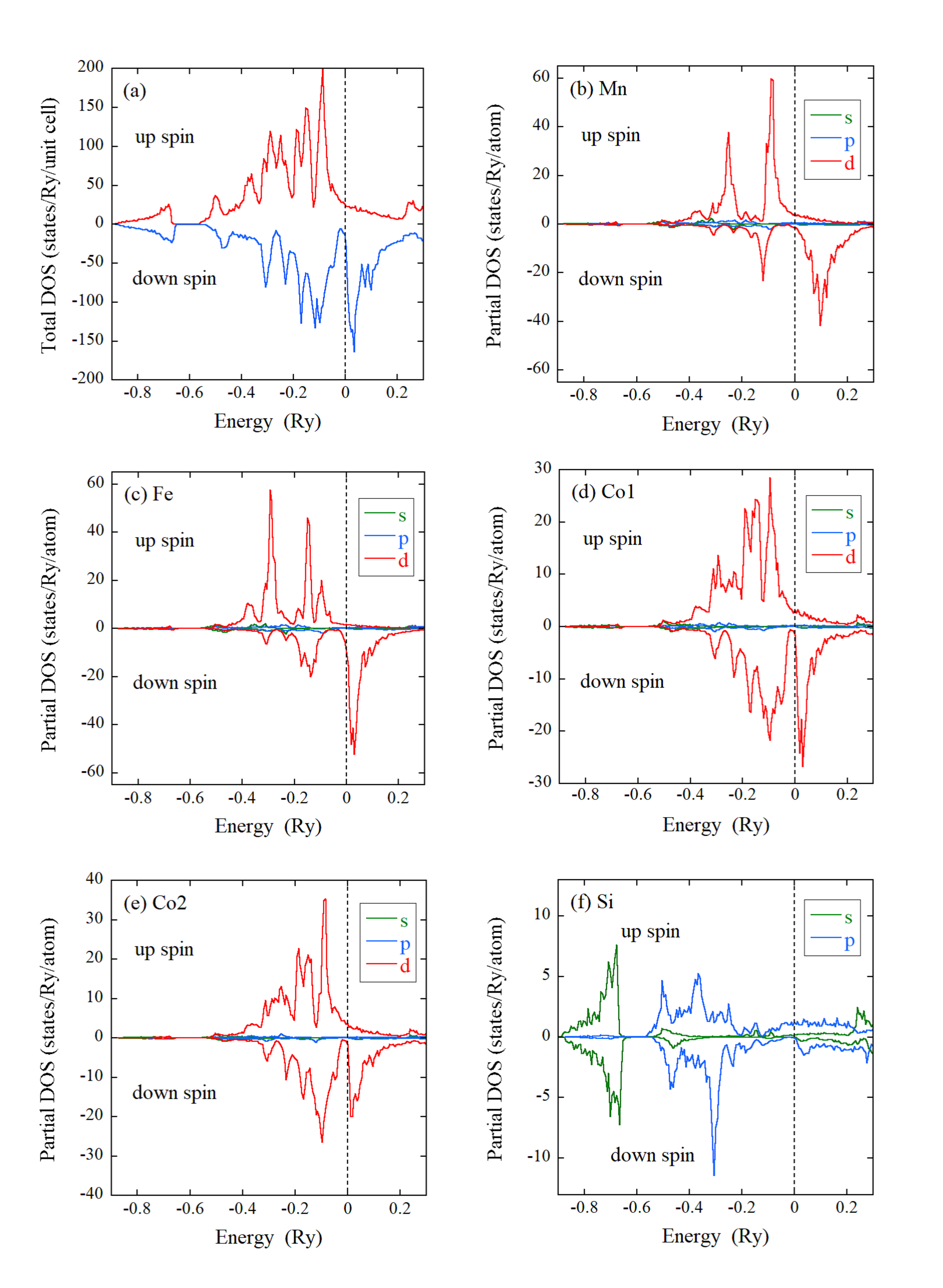}
\end{center}
\caption{(a) Electronic total density of states (DOS) of MnFeCo$_{4}$Si$_{2}$. (b) Partial DOS of Mn atom. (c) Partial DOS of Fe atom. (d) Partial DOS of Co1 atom. (e) Partial DOS of Co2 atom. (f) Partial DOS of Si atom. The Fermi energy is set to 0 Ry.}
\label{DOS}
\end{figure}

We have performed electronic-structure calculations to elucidate the ferromagnetic characteristics. 
In Figure \ref{DOS}(a), the total density of states (DOS) is plotted as a function of energy in units of Ry, with the Fermi energy set to 0 Ry. 
The different energy dependence of the total DOS between the spin-up and spin-down channels supports a ferromagnetic ground state. 
Figs.\hspace{1mm}\ref{DOS}(b)-(f) display the partial DOSs of Mn, Fe, Co1, Co2, and Si atoms, respectively. 
For Mn, Fe, Co1, and Co2, the $d$-electrons exhibit large DOS values, resulting in high magnetic moments as summarized in Table \ref{tab:moment}.
These spin moments possess positive values, indicating that Mn, Fe, and Co atoms are ferromagnetically coupled. 
Conversely, the Si atom exhibits a much smaller partial DOS, and its magnetic moment is negligible compared with those of Mn, Fe, and Co. 
The calculated $M_\mathrm{s}$ is 10.62 $\mu_\mathrm{B}$/f.u., which is in good agreement with the experimental value of 11.63 $\mu_\mathrm{B}$/f.u. 
This indicates that the magnetic structure of MnFeCo$_{4}$Si$_{2}$ is a simple ferromagnetic one, with the Mn, Fe, and Co spins aligned parallel to each other. 
Moreover, we evaluate $\mu_\mathrm{eff}$ using the magnetic spin moments of Mn, Fe, and Co atoms obtained from the electronic-structure calculations. 
$\mu_\mathrm{eff}$ is calculated by averaging the effective moments at magnetic sites as follows:
\begin{equation}
\mu_\mathrm{eff}^{2}=\frac{1}{6}\mu_\mathrm{eff-Mn}^{2}+\frac{1}{6}\mu_\mathrm{eff-Fe}^{2}+\frac{2}{6}\mu_\mathrm{eff-Co1}^{2}+\frac{2}{6}\mu_\mathrm{eff-Co2}^{2},
\label{equ:RT}
\end{equation}
where the effective moments of Mn, Fe, Co1, and Co2 are denoted as $\mu_\mathrm{eff-Mn}$, $\mu_\mathrm{eff-Fe}$, $\mu_\mathrm{eff-Co1}$, and $\mu_\mathrm{eff-Co2}$, respectively. 
The effective moment at each site is expressed as $2\sqrt{S(S+1)}$, where $S$ is half of the magnetic spin moment.
The resulting $\mu_\mathrm{eff}$ is 2.75 $\mu_\mathrm{B}$ per magnetic ion, which does not deviate significantly from the experimental value (3.29 $\mu_\mathrm{B}$).

We discuss the small discrepancy in $\mu_\mathrm{eff}$ values between calculation and experiment.
The equation $2\sqrt{S(S+1)}$, based on the localized-spin picture, is used to calculate $\mu_\mathrm{eff}$; however, this approach is inadequate for itinerant ferromagnets. 
To better assess the itinerant character, the Rhodes–Wohlfarth ratio is often used as a criterion\cite{Rhodes:PRSA1963}. 
This ratio is defined as $2S$/$M_\mathrm{s}$, where $S$ is calculated from the experimental $\mu_\mathrm{eff}$ through $\mu_\mathrm{eff}=2\sqrt{S(S+1)}$, and $M_\mathrm{s}$ is the experimental value.
In the case of an itinerant ferromagnet, the Rhodes–Wohlfarth ratio is much larger than that predicted for localized spin systems, for which $2S$/$M_\mathrm{s}$=1.
According to this criterion, when $\mu_\mathrm{eff}$ for an itinerant ferromagnet is derived from $2\sqrt{S(S+1)}$ by taking $S$ as half of the experimental $M_\mathrm{s}$, the calculated value ($\mu_\mathrm{eff}^\mathrm{local}$) is always smaller than the experimental $\mu_\mathrm{eff}$ ($\mu_\mathrm{eff}^\mathrm{exp}$).
Thus, this difference implies that the ratio $\mu_\mathrm{eff}^\mathrm{exp}/\mu_\mathrm{eff}^\mathrm{local}$ is greater than 1.
Using the literature values summarized in Ref.\cite{Takahashi:JPSJ1986}, this ratio for many itinerant ferromagnets ranges from 1.18 to 4.92, with higher values indicating a stronger itinerant character.
The deviation of the experimental $\mu_\mathrm{eff}$ from the localized spin model is attributed to spin-fluctuation effects\cite{Moriya:book}. 
For MnFeCo$_{4}$Si$_{2}$, $\mu_\mathrm{eff}^\mathrm{exp}/\mu_\mathrm{eff}^\mathrm{local}$ is 1.20, suggesting the itinerant nature likely contributes slightly to the small discrepancy between calculated and experimental $\mu_\mathrm{eff}$ values.

The value of $H_\mathrm{c}$ is not particularly large, although a higher value might be expected given the anisotropic crystal structure of MnFeCo$_{4}$Si$_{2}$. 
For example, the ferromagnetic compound Fe$_{1/4}$TaS$_{2}$, which crystallizes in a similarly anisotropic hexagonal structure, exhibits a giant $H_\mathrm{c}$ exceeding 20 kOe at 2 K\cite{Ko:PRL2011}. 
The origin of this large $H_\mathrm{c}$ has been attributed to a substantial unquenched orbital magnetic moment of $\sim$1 $\mu_{B}$ per Fe atom\cite{Ko:PRL2011}. 
In contrast to the large orbital moment reported for Fe$_{1/4}$TaS$_{2}$, the calculated orbital moments for Mn, Fe, and Co in MnFeCo$_{4}$Si$_{2}$ are two orders of magnitude smaller (see Table \ref{tab:moment}), which likely accounts for the low $H_\mathrm{c}$ observed in MnFeCo$_{4}$Si$_{2}$.

\begin{table}[t]
\caption{Spin and orbital moments of each element in MnFeCo$_{4}$Si$_{2}$ as obtained from electronic structure calculations. The total magnetic moment is 10.62  $\mu_\mathrm{B}$/f.u.}
\label{tab:moment}
\begin{tabular}{ccc}
\hline
Atom & Spin moment ($\mu_\mathrm{B}$) & Orbital moment ($\mu_\mathrm{B}$) \\
\hline
Mn & 3.18 & 0.021 \\
Fe & 2.84 & 0.061  \\
Co1 & 1.24 & 0.042 \\
Co2 & 1.15 & 0.036 \\
Si  & -0.062  & 0.0015 \\
\hline
\end{tabular}
\end{table}

\section{Summary}
This study presents an example of the experimental verification of a magnetic compound predicted by GNoME. 
We have focused on rhombohedral MnFeCo$_{4}$Si$_{2}$, which exhibits an anisotropic crystal structure in the hexagonal setting, and successfully obtained a single-phase sample adopting the predicted structure. 
The magnetic properties were characterized as those of a soft ferromagnet with $H_\mathrm{c}$= 3.7 Oe and $T_\mathrm{C}$= 1039 K.
The saturation moment of 11.63 $\mu_\mathrm{B}$/f.u. is relatively high and consistent with the value calculated using the Akai-KKR software. 
The electronic-structure calculations revealed that the magnetic species Mn, Fe, and Co exhibit parallel spin alignment, resulting in a large saturation moment. 
The small orbital moments of Mn, Fe, and Co, as estimated using the Akai-KKR software, likely account for the low $H_\mathrm{c}$ value. 
Further experimental verification of materials predicted by GNoME is necessary to assess the potential of GNoME-based materials research as a significant and reliable tool.

\section*{CRediT authorship contribution statement}
Shuhei Naganuma: Investigation. Jiro Kitagawa: Supervision, Investigation, Formal analysis, Writing - original draft, Writing - reviewing \& editing.

\section*{Declaration of competing Interest}
The authors have no conflicts to disclose.

\section*{Data availability }
Data will be made available on request.

\section*{Acknowledgments}
The authors would like to thank the support from a Grant-in-Aid for Scientific Research
(KAKENHI) (Grant No. 24K00876) and the Comprehensive Research Organization
of Fukuoka Institute of Technology.


\begin{thebibliography}{99}
\bibitem{Stanev:CM2021}
V. Stanev, K. Choudhary, A.G. Kusne, J. Paglione, and I. Takeuchi: Commun Mater. {\bf 2}, 105 (2021). 

\bibitem{Choudhary:NCM2022}
K. Choudhary, B. DeCost, C. Chen, A. Jain, F. Tavazza, R. Cohn, C.W. Park, A. Choudhary, A. Agrawal, S.J.L. Billinge, E. Holm, S.P. Ong, and C. Wolverton: npj Comput. Mater. {\bf 8}, 59 (2022).

\bibitem{Castro:NAM:2020}
P.B.d. Castro, K. Terashima, T.D. Yamamoto, Z. Hou, S. Iwasaki, R. Matsumoto, S. Adachi, Y. Saito, P. Song, H. Takeya, and Y. Takano: NPG Asia Mater. {\bf 12}, 35 (2020). 

\bibitem{Rao:Science2022}
Z. Rao, P.-Y. Tung, R. Xie, Y. Wei, H. Zhang, A. Ferrari, T.P.C. Klaver, F. K\"{o}rmann, P. Thoudden Sukumar, A. Kwiatkowski da Silva, Y. Chen, Z. Li, D. Ponge, J. Neugebauer, O. Gutfleisch, S. Bauer, and D. Raabe: Science {\bf 378}, 6615 (2022). 

\bibitem{Hou:AAMI2019}
Z. Hou, Y. Takagiwa, Y. Shinohara, Y. Xu, and K. Tsuda: ACS Appl. Mater. Interfaces {\bf 11}, 11545 (2019).

\bibitem{Jiang:JMST2022}
L. Jiang, C. Wang, H. Fu, J. Shen, Z. Zhang, and J. Xie: J. Mater. Sci. Technol. {\bf 98}, 33 (2022).

\bibitem{Merchant:Nature2023}
A. Merchant, S. Batzner, S.S. Schoenholz, M. Aykol, G. Cheon, and E.D. Cubuk: Nature {\bf 624}, 80 (2023).

\bibitem{Jain:APLM2013}
A. Jain, S.P. Ong, G. Hautier, W. Chen, W. D. Richards, S. Dacek, S. Cholia, D. Gunter, D. Skinner, G. Ceder, and K.A. Persson: APL Mater. {\bf 1}, 011002 (2013).

\bibitem{Kitagawa:JMMM2018}
J. Kitagawa and K. Sakaguchi: J. Magn. Magn. Mater. {\bf 468}, 115 (2018).

\bibitem{Kitagawa:MRX2023}
J. Kitagawa, H. Nomura, and T. Nishizaki: Mater. Res. Express {\bf 10}, 106102 (2023).

\bibitem{Kitagawa:JMMM2022}
J. Kitagawa: J. Magn. Magn. Mater. {\bf 563}, 170024 (2022).
\bibitem{Akai:JPSJ1982}
H. Akai: J. Phys. Soc. Jpn. {\bf 51}, 468 (1982). 

\bibitem{Izumi:SSP2007}
F. Izumi and K. Momma: Solid State Phenom. {\bf 130}, 15 (2007). 

\bibitem{Tsubota:SR2017}
M. Tsubota and J. Kitagawa: Sci. Rep. {\bf 7}, 15381 (2017).

\bibitem{Mohamad:JNST2018}
A. Mohamad, Y. , H. Muta, K. Kurosaki and S. Yamanaka: Journal of Nuclear Science and Technology {\bf 55}, 1141 (2018).

\bibitem{Kitagawa:JALCOM2022}
J. Kitagawa, K. Hoshi, Y. Kawasaki, R. Koga, Y. Mizuguchi and T. Nishizaki: J. Alloys Compd. {\bf 924}, 166473 (2022).

\bibitem{Sarr:JALCOM2025}
I. Sarr, O. Isnard, A. Verniere and L.V.B. Diop: J. Alloys Compd. {\bf 1037}, 182354 (2022).

\bibitem{Oikawa:APL2001}
K. Oikawa, L. Wulff, T. Iijima, F. Gejima, T. Ohmori, A. Fujita, K. Fukamichi, R. Kainuma, and K. Ishida: Appl. Phys. Lett. {\bf 79}, 3290 (2001).

\bibitem{Yu:APL2003}
M.-H. Yu, L.H. Lewis, and A.R. Moodenbaugh: J. Appl. Phys. {\bf 93}, 10128 (2003).

\bibitem{Rhodes:PRSA1963}
P. Rhodes and E.P. Wohlfarth: Proc. Roy. Soc. A {\bf 273}, 247 (1963).

\bibitem{Takahashi:JPSJ1986}
Y. Takahashi: J. Phys. Soc. Jpn. {\bf55}, 3553 (1986).

\bibitem{Moriya:book}
T. Moriya: {\it Spin fluctuations in itinerant electron magnetism} (Springer, Berlin) 2012.

\bibitem{Ko:PRL2011}
K.-T. Ko, K. Kim, S.B. Kim, H.-D. Kim, J.-Y. Kim, B.I. Min, J.-H. Park, F.-H. Chang, H.-J. Lin, A. Tanaka, S.-W. Cheong: Phys. Rev. Letts. {\bf 107}, 247201 (2011).

\end{thebibliography}
\end{document}